\documentstyle[preprint,aps,epsfig]{revtex}
\begin{document}
\draft
\title{Theory of ac electrokinetic behavior of spheroidal cell suspensions with an intrinsic dispersion}
\author{Lei Gao$^{1,2}$, J. P. Huang$^1$ and K. W. Yu$^1$}
\address{$^1$ Department of Physics, The Chinese University of Hong Kong, Shatin,  N.T.,  Hong Kong, China}
\address{$^2$ Department of Physics, Suzhou University, Suzhou 215006, China}
\maketitle

\begin{abstract}

The dielectric dispersion, dielectrophoretic (DEP) and electrorotational (ER) 
spectra of spheroidal biological cell suspensions
with an intrinsic dispersion in the constituent dielectric constants 
are investigated. By means of the spectral representation method, we  
express analytically the characteristic frequencies and dispersion 
strengths  both for the effective dielectric constant and 
the Clausius-Mossotti factor (CMF). We identify four and six 
characteristic frequencies for the effective dielectric spectra and CMF 
respectively, all of them being dependent on the depolarization factor (or the cell shape). 
The analytical results allow us to 
examine the effects of 
the cell shape, the dispersion strength and the intrinsic  frequency  on the dielectric dispersion,
 DEP  and ER spectra. 
Furthermore, we include the local-field effects due to the mutual interactions between cells in a dense 
suspension, and study
the dependence of co-field or anti-field dispersion peaks on the volume fractions.

\end{abstract}

\vskip 5mm 

\newpage
\section{Introduction}

The polarization of biological cells has a wide range of practical applications like 
manipulation, trapping or separation of biological cells~\cite{Gimsa,Gimsa99}.
Dielectric spectroscopy~\cite{Asami80}, dielectrophoresis (DEP)~\cite{Fuhr}  and 
electrorotation (ER)~\cite{Gimsa91} offer a unique capability of 
monitoring the dielectric properties of dispersions of colloids and 
biological cells. 
Under the action of external fields, these particles exhibit rich 
fluid-dynamic behaviors as well as various dielectric responses. 
Hence, it is of importance to investigate their frequency-dependent responses 
to ac electric fields, which yields valuable information on the 
structural (Maxwell-Wagner) polarization effects. 
The polarization is characterized by a variety of characteristic 
frequency-dependent changes known as the dielectric dispersion, whose 
spectra are helpful to analyze the inhomogeneous systems, including biological cell 
suspensions and tissues~\cite{Asami2}.  Many factors exert influences on the effective dielectric behavior of the 
system such as  orientation of dipoles, surface conductance and the cell shape~\cite{Lei}.
However, some factors can be dominant at certain ranges of frequencies (for example, the experimental data revealed 
that the low frequency subdispersions were dependent on the cell shape~\cite{Asami3}).

The DEP is used to describe the motion of the particles 
caused by the dielectric polarization effects in non-uniform electric 
field~\cite{Pohl}. The DEP force drives the particles towards high 
intensity (positive 
DEP) or towards regions of minimal field intensity (negative DEP). Because 
of the interaction (i.e., DEP force) between the induced dipole and the external 
electric field, the particles can be levitated in the medium~\cite{Tombs}.

The ER behavior is caused by the existence of a phase 
difference between the field-induced dipole moment and the external 
rotating field, which results in a  torque to cause  the particle to rotate. 
In the dilute limit, the ER of individual cell can be predicted by 
ignoring the mutual interaction between the cells, and hence may be considered as
an isolated particle in rotation~\cite{JPCM}. 

In this paper, we will apply the spectral representation theory~\cite{DJ} to investigate the ac electrokinetic behavior
including the dielectric dispersion, 
DEP and ER spectra of biological cell suspensions.
The object of the present investigation is threefold. First, it is instructive to 
consider the effect of  the intrinsic dielectric dispersion.
 Actually, such an intrinsic dielectric dispersion
often occurs due to the surface conductivity~\cite{Schwarz,Gao2} or the inhomogeneous structure
such as  the coated shells of biological cells or 
colloidal suspensions~\cite{Wang1}.  Secondly, many cells exist in the form of the nonspherical shape such as
the fission yeast cell~\cite{Asami3} and the red blood cell~\cite{Miller}, and hence 
the effect of cell 
shape needs to be considered~\cite{Paul1,Paul2}. Thirdly, the volume fractions of the biological cell suspensions are 
generally not in the dilute limit and can even exceed  $0.1$~\cite{Asami2,Asami3,Bordi}, thus 
the local field effect due to the average electrostatic interaction between cells must be taken into account~\cite{PRE2}.
In view of this, the effective dielectric
dispersion spectra will be  studied based on
 Maxwell-Garnett type approximation~\cite{Gao1}, which 
involves an exact calculation of the field induced in the host medium by  a single ellipsoidal cell and an approximate treatment of its distortion by the electrostatic interaction between different cell suspensions.
 Moreover, the Clausius-Mossotti factor (CMF), 
which determines the DEP and ER spectra, will be 
modified by replacing the dielectric constant of the host medium with the effective one.

Being beyond this work, we noticed an alternative paper~\cite{Foster}, in which the authors considered the electrorotation and levitation of spherical cells or colloidal particles 
with/without the membrane-covered shells in the dilute limit. However, our investigation can be valid not only for nonspherical cell suspensions without membrane-covered shells, but also for non-dilute limit suspensions. Furthermore,  we adopt the 
spectral representation method~\cite{DJ}, 
which offers the advantage of the separation of material 
parameters from the 
geometric information, to simplify the derivation of the analytical expressions for the characteristic  frequencies and dispersion strengths 
of the effective dielectric permittivity and the CMF, respectively.

\section{Formalism}

We consider a composite system in which biological cells of dielectric constant 
$\tilde{\epsilon}_1$
with the volume fraction $p$ are dispersed in an isotropic host medium with dilectric constant 
$\tilde{\epsilon}_2=\epsilon_2+\sigma_2/(j 2\pi f)$ with $j=\sqrt{-1}$,
 $f$ being 
the frequency of the applied field.  $\tilde{\epsilon}_1$ is 
assumed to exhibit an intrinsic dielectric dispersion, i.e.,
\begin{equation}
\tilde{\epsilon}_1=\epsilon_1+\frac{\triangle\epsilon_1}{1+j f/f_1}+\frac{\sigma_1}{j 2\pi f},
\end{equation}
where $\epsilon_1$ ($\sigma_1$) is the limiting high-frequency 
(low-frequency) dielectric constant (conductivity),
$\triangle \epsilon_1$ represents the dielectric dispersion strength with  a characteristic 
frequency $f_1$.  Biological cells are further assumed to be spheroidal in shape. 
Such an assumption is supported by 
Bohren {\it et al}~\cite{Bohren} who suggested that dielectric dispersion 
spectra of particles of arbitrary shape can  be approximated by 
these of spheroidal particles.  

Then we will adopt the spectral representation to
investigate the ac electrokinetic behavior such as the dielectric dispersion, 
dielectrophoretic and electrorotational spectra
of an inhomogeneous system in which  spheroidal biological cells are randomly distributed.

\subsection{ Dielectric dispersion spectrum}

Generally, for cell suspensions of arbitrary shape, the spectral representation can only be
solved numerically. However, 
for randomly-oriented spheroidal cell suspensions, within the mean-field theory, 
the effective complex dielectric constant $\tilde{\epsilon}_e$ can 
be written as~\cite{Gao1}
\begin{equation}
\tilde{\epsilon}_e=\tilde{\epsilon}_2\frac{1+p\left[b_z(1-L_z)+2b_{xy}(1-L_{xy})\right]}
{1-p(b_zL_z+2b_{xy}L_{xy})},
\end{equation}
where $b_{k}\equiv(\tilde{\epsilon}_1-\tilde{\epsilon}_2)/\{3\left[\tilde{\epsilon}_2+
L_k(\tilde{\epsilon}_1-\tilde{\epsilon}_2)\right]\}$ $(k=z,xy)$ is called the Clausius-Mossotti factor (CMF), 
and $L_z$ 
[$L_{xy}=(1-L_z)/2$] are 
the depolarization factors along the $z$ ($x$ or $y$) axis of the spheroids. These depolarization factors depend on the aspect ratio $q\equiv c/a$~\cite{Gao1}, where $a$ ($=b$), $c$ are the semi-axes of 
a spheroid along the Cartesian axes. For the prolate-spheroidal cell ($q>1$), we have $0<L_z<1/3$; while for the oblate-spheroidal one ($q<1$), we have $1/3<L_z<1$. 
  Actually, once a $q$ is given, $L_z$ can be obtained uniquely and thus  used to 
 indicate the shape of spheroidal particles.

By invoking the spectral representation and introducing the dimensionless parameter
$\tilde{s}\equiv \tilde{\epsilon}_2/(\tilde{\epsilon}_2-\tilde{\epsilon}_1)$, we 
rewrite Eq.(2) as 
\begin{equation}
\tilde{\epsilon}_e=\tilde{\epsilon}_2\left(1-\frac{W_1}{\tilde{s}-x_1}
-\frac{W_2}{\tilde{s}-x_2}\right),
\end{equation}
where the poles $x_1$ and $x_2$ are given as
\begin{equation}
x_{1,2}=\frac{1}{12}\left[3-2p+3L_z \pm\sqrt{(3-2p+3L_z)^2-72(1-p)L_z(1-L_z)}\right],
\end{equation}
Correspondingly, the residues $W_1$ and $W_2$ have 
\begin{equation}
W_1=\frac{p(1+3L_z-6x_1)}{6(x_2-x_1)},\,\, W_2=\frac{p(1+3L_z-6x_2)}{6(x_1-x_2)}.
\end{equation}
It is easy to check that $W_1$ and $W_2$ satisfy the sum rule $W_1+W_2=p$.

In order to obtain the analytic expressions for the 
dielectric dispersion strengths and the characteristic 
frequencies of the dielectric dispersion spectra, we  then introduce
two contrast parameters $s=\epsilon_2/(\epsilon_2-\epsilon_1)$ and 
$t=\sigma_2/(\sigma_2-\sigma_1)$, firstly defined by Lei {\it et al}~\cite{Lei}, 
and derive an equality for biological 
cells possessing an intrinsic 
dispersion, 
\begin{equation}
\frac{1}{\tilde{s}-x}=\frac{1}{s-x}+\frac{A(x)}{1+jf/f_{c1}(x)}+\frac{B(x)}{1+jf/f_{c2}(x)},
\end{equation}
where
\begin{equation}
f_{c1,c2}(x)=\frac{f_1f_2+f_2f_3+f_3f_1\mp\sqrt{(f_1f_2+f_2f_3+f_3f_1)^2-4f_1f_2f_3^2}}{2 f_3},
\end{equation}
and
\begin{equation}
A(x)=\left[
\frac{s-t}{(s-x)(t-x)}\cdot\left(\frac{f_1-f_{c1}}{f_1f_{c1}}\right) 
+\frac{s}{(s-x)x}\cdot\frac{1}{f_3}
\right] \frac{f_1f_2}{f_{c2}(x)-f_{c1}(x)},
\end{equation}

\begin{equation}
B(x)=\left[
\frac{s-t}{(s-x)(t-x)}\cdot\left(\frac{f_{c2}-f_1}{f_1f_{c2}}\right)
-\frac{s}{(s-x)x}\cdot \frac{1}{f_3}
\right]\frac{f_1f_2}{f_{c2}(x)-f_{c1}(x)},
\end{equation}
with
\begin{equation}
f_2(x)=\frac{1}{2\pi}\cdot\frac{s\sigma_2(t-x)}{\epsilon_2 t(s-x)},\,\, f_3(x)=\frac{1}{2\pi}
\cdot\frac{\sigma_2(t-x)}{x\triangle \epsilon_1 t}.
\end{equation}

After simple manipulations, we can express Eq.(3) as,
\begin{equation}
\tilde{\epsilon_e}=\epsilon_H+\sum_{i=1}^4 
\frac{\triangle\epsilon_{ei}}{1+j (f/f_{ei})}+\frac{\sigma_L}{j 2\pi f}\equiv \epsilon_e+\frac{\sigma_e}{j 2\pi f},
\end{equation}
with
\begin{equation}
\epsilon_H=\epsilon_2\left(1-\frac{W_1}{s-x_1}-\frac{W_2}{s-x_2}\right), \quad \sigma_L=\sigma_2\left(1-\frac{W_1}{t-x_1}-\frac{W_2}{t-x_2}\right)
\end{equation}
and 

\begin{eqnarray}
\triangle\epsilon_{e1}=W_1A(x_1)\left(\frac{\sigma_2}{2\pi f_{c1}(x_1)}-\epsilon_2\right),\,\, f_{e1}=f_{c1}(x_1),\\
\triangle\epsilon_{e2}=W_1B(x_1)\left(\frac{\sigma_2}{2\pi f_{c2}(x_1)}-\epsilon_2\right),\,\, f_{e2}=f_{c2}(x_1).
\end{eqnarray}
Similarly, $\triangle\epsilon_{ei}$ and $f_{ei}$ for $i=3,4$ can be easily got by our replacing $W_1$ and $x_1$ with $W_2$ and $x_2$ from above two equations.

Thus, within the spectral representation, for spheroidal  suspensions with one intrinsic 
dispersion, we successfully
obtain four characteristic
 freqencies $f_{ei}$ $(i=1,4)$ with the 
 dispersion strengths $\triangle \epsilon_{ei}$ in terms of the geometric parameters  ($L_z$ and $p$)  
as well as physical parameters ($s$, $t$, $\triangle\epsilon_1$,
$\sigma_2$ and $\epsilon_2$).  
 
 In Fig.1, we numerically calculate four characteristic frequencies and dispersion strengths against $L_z$. As is evident from the figure, the characteristic freqencies and the corresponding  dispersion 
 strengths are strongly dependent on the shape of cell suspensions
 and exhibit nonmonotonical behavior with 
 increasing $L_z$.  For these four characteristic frequencies, two of them are located
  at  frequencies higher than
 $10^6Hz$, while another two are at  frequencies lower than $10^6Hz$.  Such a separation results from the 
 fact that we take into account one intrinsic dispersion in the dielectric cell suspensions. 
 Furthermore, due to the nonspherical shape of cell suspensions, there are two characteristic frequencies in the high-frequency (or low-frequency) region. When the shape is far from spherical (say
 $L_z\rightarrow0 $ or $1$), the difference between these two frequencies becomes large.
 Conversely,  for spherical inclusions ($L_z=1/3$),  only two dispersion strengths are nonzero, 
 leading to two-main steps in the dielectric dispersion spectra, as will be shown in the next section.

\subsection{ dielectrophoretic  and electrorotational  spectra}

The spectral form of Eq.(6) also allows us to investigate the dielectrophoretic  and electrorotational 
 behavior.

It is known that the time average DEP force $F(f)$ and ER torque $\Gamma(f)$ excerted on the spheroidal cell 
 suspensions with $z$-orientation parallel to the electric field are expressed as~\cite{Cruz,Wang2}
 \begin{equation}
 F(f)=1.5v\tilde{\epsilon}_2{\it Re}[b_z(\tilde{\epsilon}_1,\tilde{\epsilon}_2)]|\bigtriangledown {\bf E}^2_{(rms)}|,\,\, \Gamma(f)=-3v\tilde{\epsilon}_2{\it Im}[b_z(\tilde{\epsilon}_1,\tilde{\epsilon}_2)]{\bf E}_0^2,
 \end{equation}
 where $v$ is the volume of spheroidal cell, ${\bf E}_{(rms)}$ is the root-mean 
 square magnitude of the imposed ac electric
 field and ${\it Re}[\cdot\cdot\cdot]$, ${\it Im}[\cdot\cdot\cdot]$ represent the real and imaginary parts of  $b_z$, which can be written under the spectral representation as 
 \begin{equation}
 b_z(\tilde{\epsilon}_1,\tilde{\epsilon}_2)=\frac{1}{3}\cdot
 \frac{\tilde{\epsilon}_1-\tilde{\epsilon}_2}
 {\tilde{\epsilon}_2+L_z(\tilde{\epsilon}_1-\tilde{\epsilon}_2)}
 =-\frac{1}{3}\cdot \frac{1}{\tilde{s}-L_z} 
\end{equation}
Eq.(16) is independent of the volume fraction $p$ and thus 
 is valid only in the dilute limit~\cite{Foster}. However, for non-dilute volume fractions, we must 
consider the 
 local field effect due to mutual interaction between the spheroidal cells
  and modify Eq.(16) as,
 \begin{equation}
 b_z(\tilde{\epsilon}_1,\tilde{\epsilon}_2)=\frac{1}{3}\cdot\frac{\tilde{\epsilon}_1
 -\tilde{\epsilon}_e}{\tilde{\epsilon}_e+L_z(\tilde{\epsilon}_1-\tilde{\epsilon}_e)}.
 \end{equation}
 
 Substituting Eq.(2) into (17), we have
 \begin{eqnarray}
 b_z&=&\frac{2(p-1)\tilde{s}^2+(1-p)(1+L_z)\tilde{s}-(1-p)L_z(1-L_z)}
 {6\tilde{s}^3-(3+4p+9L_z-6pL_z)\tilde{s}^2+(p+6L_z-3pL_z)\tilde{s}+3(p-1)L_z^2(1-L_z)} 
 \nonumber\\
 &=&\frac{V_1}{\tilde{s}-s_{1}}+
 \frac{V_2}{\tilde{s}-s_{2}}+\frac{V_3}{\tilde{s}-s_{3}},
 \end{eqnarray}
 where $s_{i}$ $(i=1,2,3)$ are three roots of polynomial equations
 \begin{equation}
 y^3-\frac{3+4p+9L_z-6pL_z}{6}y^2+\frac{p+6L_z-3pL_z}{6}y+\frac{(p-1)L_z^2(1-L_z)}{2}=0,
 \end{equation}
while the residues $V_1$, $V_{2}$ and $V_{3}$ can be obtained from
 \begin{eqnarray}
 V_{1}+V_{2}+V_{3}&=&\frac{1}{3}(p-1) \nonumber \\
 V_{1}(s_{2}+s_{3})+V_{2}(s_{3}+s_{1})+V_{3}(s_{1}+s_{2})&=&\frac{1}{6}(p-1)(1+L_z) \\
 V_{1}s_{2}s_{3}+V_{2}s_{1}s_{3}+V_{3}s_{1}s_{2}&=&\frac{1}{6}(p-1)L_z(1-L_z) \nonumber.
 \end{eqnarray}
 Note that these residues  again satisfy the sum rule 
 $V_1+V_2+V_3=-(1-p)/3$.
 
 Introducing Eq.(6) into Eq.(18) leads to
 \begin{equation}
 b_z=\sum_{i=1}^3\left[\frac{V_i}{s-s_i}+\frac{V_iA(s_i)}{1+jf/f_{c1}(s_i)}+
 \frac{V_iB(s_i)}{1+jf/f_{c2}(s_i)}\right].
 \end{equation}
 
 The normalized DEP  force and normalized ER torque can be defined as 
 \begin{equation}
 NDEP = {\it Re}\left[b_z(\tilde{\epsilon}_1,\tilde{\epsilon}_2)\right] =
 \sum_{i=1}^3\left[\frac{V_i}{s-s_i}+\frac{V_iA(s_i)}{1+[f/f_{c1}(s_i)]^2}+
 \frac{V_iB(s_i)}{1+[f/f_{c2}(s_i)]^2}\right],
 \end{equation}

 and 
 \begin{equation}
 NER=-{\it Im}\left[b_z(\tilde{\epsilon}_1,\tilde{\epsilon}_2)\right]
 =
\sum_{i=1}^3
 \left[\frac{V_iA(s_i)f/f_{c1}(s_i)}{1+[f/f_{c1}(s_i)]^2}+
 \frac{V_iB(s_i)f/f_{c2}(s_i)}{1+[f/f_{c2}(s_i)]^2}\right].
 \end{equation}
 
 Based upon the  spectral representation,  the  expressions for the dispersion 
 strengths and characteristic frequencies of 
CMF are also explicitly obtained.  
Six CMF characteristic frequencies   and  dispersion 
 strengths  against $L_z$ are shown in Fig.2.
                                 
 Again, $L_z$ plays an important role in determining the CMF
 characteristic frequencies and the corresponding 
  strengths. Within our model, six CMF characteristic frequencies are predicted; three of them
  are located below  $10^6Hz$ and another three do above $10^6Hz$, similar as that observed in Fig.1.
  Due to this, the DEP (or ER) spectra
  are maining characterized by two-step rapid changes (or two dominant peaks).  
  For three typical cases, we also find, six CMF strengths 
  are all positive in Fig.2(a); all negative in Fig.2(b) and three strengths are negative and three are positive in
  Fig.2(c). such behavior 
  will result in co-field or anti-field rotation of the particle for ER spectra. 
  We remark that as both the intrinsic dispersion and local-field effects are considered 
   simutaneously, we predict four characteristic frequecies, characterized by four non-zero dispersion strengths even for $L_z=1/3$.

  \section{Numerical results}
  
  We are now in a position to calculate the dielectric dispersion, and DEP and ER spectra
  based on the model put forward in the previous section.

  In Fig.3-5, we investigate the effect of depolarization factor $L_z$ 
  on the dielectric response [denoted by 
  $\epsilon_e/\epsilon_0$ and $\sigma_e$ defined in Eq.(11)], DEP and ER spectra using Eq.(22) and Eq.(23).
  
  The dielectric response mainly exhibits  two sub-dispersions characterized by two-step rapid decrease in
   Fig.3(a), Fig.4(a) and Fig.5(a). It is evident that the influence of $L_z$ (the shape)  on 
   the effective dielectric constant (or the effective conductivitty) is significant 
   in the low-frequency region where  $f<10^4Hz$ (or in the high-frequency region where
   $f>10^8 Hz$).  This is in agreement with previous experimental conclusions that 
   the low-frequency subdispersion for $\epsilon_e$ depends on the cell shape, whereas the high-frequency 
   is independent of it~\cite{Asami3}. The dielectric permittivity for oblate spheroidal particles, say $L_z=0.95$, 
   in the low-frequency region  can be as four times as the one for  spherical inclusions. Thus, by taking 
   into account the shape of cell suspensions, it is possible to reduce some discrepancies between 
   the theoretical and experimental results on the dielectric dispersion spectra~\cite{Asami3,Bordi}. 
   
      For DEP force, it exhibits strong sensitivity to the depolarization factor (the shape).
    When DEP force is positive (the particles will be attracted towards
   the field-generating electrodes), with increasing $L_z$,  such an
    attractive force becomes weak. 
   However, when the DEP force is negative (the particles will be repelled from the electrodes), the larger $L_z$, the stronger repulsion will be.
    Furthermore, we predict one or two shape-dependent crossover 
   frequencies at which there is no net force on the cell particle.  
   The  crossover frequency is a monotonically increasing or decreasing function of $L_z$, 
   dependent on  whether the variation of DEP force around it is
   negative or positive [see the inserts in Fig.3(b), Fig.4(b) and Fig.5(a)].
   This is an interesting result. To the best of our knowledge, the dependence of crossover frequency on the
   spheroidal shape is  reported here for the first time. 
     
  For ER spectra, they exhibit two ER characteristic frequencies with both positive strengths in Fig.3(b),
   both negative strengths in Fig.4(b), and negative and positive strengths in Fig.5(b) (This 
   has been observed in experimental electrorotational 
   spectra of the yeast cell~\cite{Zhou,Kriegmaier}). 
    Positive ER torque means the co-field rotation of the particle. In this case, 
   with increasing $L_z$, the ER characteristic frequency for ER will be blue-shifted accompanied with the decrease of the 
   dispersion strengths. However, when  the ER torque is negative (the particle exhibits the
   anti-field
      rotation), both the ER characteristic frequencies and the strengths decrease concomitantly with $L_z$.
   Moreover, 
   we find that the variation of the geometric parameter $L_z$ does not change the polarity of ER peak.

   We emphasize to point out that, although we analytically predict four and six characteristic frequencies for 
   dielectric dispersion and ER spectra, these frequencies are mainly 
   located in two regions or some of  their dispersion strengths are quite small in comparison with others,
resulting in only two effective characteristic frequencies at which 
   the rate of change of the dielectric permittivity (or DEP force), and  the rotational 
   torque peak attain their maxima.  This can be well understood from Fig.6. 
   
   As we have included the dielectric dispersion form in $\tilde{\epsilon}_1$, 
it would be interesting to investigate how the dispersion strength $\triangle \epsilon_1$ 
   and  the intrinsic 
    frequency $f_1$ affect the dielectric dispersion and ER spectra.
   
   In Fig.7, we examine the effect of $\triangle \epsilon_1$ on the 
   dielectric response (the left panel) and ER spectra (the right panel).
   Clearly, one common feature of all these dielectric spectra is that increasing $\triangle \epsilon_1$ yields
   increasing dielectric permittivity in the low-frequency region.
   As previous models do negelect the intrinsic dispersion effect, 
   the theoretical results in the low-frequency region are less than experimental reports~\cite{Asami3,Bordi}. 
   We think it 
   would be helpful to give the closest fit with experimental data by suitable adjustment of 
   both $\triangle \epsilon_1$ and $L_z$. For ER spectra, when $\triangle \epsilon_1=0$, only one dominant 
   peak is predicted as expected. For a finite
   $\triangle \epsilon_1$, two dominant peaks at low and high characteristic frequencies  arise. 
   With increasing $\triangle \epsilon_1$, the characteristic low frequency shifts towards a long wavelength;
   while the characteristic high frequency shifts towards a short one, leading to 
   further separation between two characteristic frequencies for larger $\triangle\epsilon_1$.
   At the same time, the adjustment of $\triangle \epsilon_1$ can also change the peak value much or less and even dominate the co-field or 
   anti-field rotation through changing the polarity of the peak (see the upper right column).

    In Fig.8, we plot the dielectric dispersion and ER spectra versus the frequency $f$ for 
   different $f_1$. 
   It is evident that, $f_1$  mainly play a role in  the characteristic 
   high frequencies for both the effective dielectric dispersion and ER spectra. 
   Generally speaking, the characteristic high frequency increases 
   with increasing $f_1$, in accord with the analytical formula  from Eq.(7). Interestingly, 
   the increase in $f_1$ can also lead to the change of the polarity of the ER peak in the high-frequency region 
   (see the middle right pannel).
   
   In order to take into account the mutual interaction bewtween the suspended particles, we adopt 
   Eq.(17) instead of Eq.(16). When the mutual interaction is taken into account, the
    ER peak must be reduced, as expected in Fig.9. 
   With further increasing $p$, the mutual interaction becomes strong, leading to 
   the serious depression of ER peak and the slight blue-shift (or red-shift) in  the 
   characteristic low frequency (or high frequency). These results are also new.
   
   \section{discussion and conclusion}
   
   In this work, we have investigated both analytically and numerically the ac electrokinetic behavior, i.e.,
the dielectric dispersion behavior, the DEP and ER 
 spectra of nonspherical cell suspensions with an intrinsic dispersion based on 
  the spectral representation method. The dependence of the ac electrokinetic behavior on 
the depolarization factor is studied in detail.
   We find that an intrinsic dispersion in the dielectric constant of spheroidal cells
    can lead to four and six dispersion in the effective 
   dielectric constant and the CMF respectively.
   We 
 also  find that both the intrinsic dispersion strength $\triangle \epsilon_1$ and 
   the characteristic  frequency $f_1$ can change the polarity of ER torque and thereby 
   causing
   the co-field or anti-field rotation. Furthermore, 
   by taking into account the local-field effect
   from the mutual interaction, we examine the ER spectra for various volume fractions, and find that increasing 
   volume fractions can result in the decrease of both the strengths and the difference between two
   characteristic low and high frequencies indeed. Thus it is possible to obtain good agreement between theoretical 
   predictions and experimental data by the suitable adjustment of both the geometric parameters (for example, the 
   particle shape) and the physical parameters (for example, the dispersion strength and 
   the characteristic frequency of cell suspensions). 
   On the contrary, such a fitting is very useful to obtain 
   the relevant physical information of cell suspensions.
  
   Here a few comments are in order.  We obtain 
   four and six characteristic frequencies for the effective dielectric permittivity and 
   CMF. Generally, none of the characteristic frequencies for the dielectric spectra is 
   equal to those of CMF, as the poles $x_1$ and $x_2$ [Eq.(4)] are quite different from 
   the poles $s_1$, $s_2$ and $s_3$ of CMF. In the dilute limit and for spherical inclusions, both 
   Eq.(3) and Eq.(17) yield two characteristic frequencies;  the smaller $p$, the less differences 
   between these frequencies~\cite{Wang1} are. 
   Numerically, the dielectric dispersion, DEP and ER spectra mainly exhibit 
   two rapid changes or two dominant peaks.  Here, we also show
   a four-step rapid decrease in dielectric dispersion spectra (see Fig.10).
   However, due to  small  differences between the CMF frequencies or small strengths of some 
   frequencies in comparison with others, it is difficult to show all six characteristic 
   frequencies in ER spectra clearly.  A possible way to achieve this is to consider cell suspensions with
   large volume fractions while being far from the spherical shape.

    We demonstrate theoretically that the shape effect on the effective dielectric constant
    is significant in the low-frequency region, in accord with experimental conclusions~\cite{Asami3}. 
   In  previous work, one of the authors 
   have presented a first-principle study of the dielectric dispersion of 
fission yeast cell suspensions~\cite{Lei}. 
As the  derivation is based on the assumption that cell suspensions are in the 
dilute limit and do not exhibit an  intrinsic dispersion, the predicted 
two characteristic frequencies
 are independent of the volume fractions and some discrepancies between the theory and experiment on fission yeast 
 cells still exist. We believe that our present  
formula can be applied also  and  the discrepancies between theoretical and experimental results may  
further be reduced.

  For shelled spheroidal particles dispersed in the host medium, we have found one peak in the electrorotation
  assay, under the assumption that the ratio of shell to the host dielectric constant is real~\cite{JPCM}. 
  The theory yields good agreement with the experimental results only in the high-frequency region.
 In fact, for such kind of three component system, the effective dielectric response $\epsilon_{cs}$ for the 
 coated shells can 
  be firstly obtained in a self-consistent way~\cite{Gao1} and will exhibit at least one dispersion.
  The three-component composites can then  be equivalent to the solid spheroidal particles 
  with $\epsilon_{cs}$ including 
  one dispersion term suspended in the host medium. Thus our present model can be safely used, 
   and the 
  co-field rotational peak in the low-frequency rigion may also be predicted. Finally, we should remark that, for nonspherical 
cells, the intrinsic dispersion may depend on the orientation of a spheroid, which will yield dielectric anisotropy in cell suspensions.  In this work, we take into account 
the geometric anisotropy as well as the dielectric isotropy. The work including both geometric and dielectric anisotropies is in progress and will be reported elsewhere.

 \section*{Acknowledgments}
This work was supported by the RGC Earmarked Grant under 
project numbers CUHK 4245/01P. L. G. acknowledges the financial supports of  the National Natural Science Foundation of China  under Grant No.10204017 and   of the Science Foundation of Jiangsu Province  under Grant No. BK2002038. K.W.Y. thanks Prof. A. R. Day
for his interest in our work and his suggestion to examine the effects of inhomogeneities
on the dielectric response of biological cell suspensions. 


\newpage

\begin{figure}[h]
\caption{The characteristic frequencies and the dispersion strengths against the depolarization factor $L_z$ 
for $p=0.2$ and (a):  $s=1.625$, $t=-0.053$, $\epsilon_2=78\epsilon_0$, $\sigma_2=10^{-5}S/m$, 
$\triangle \epsilon_1=200\epsilon_0$ and $f_1=1.59\times 10^{7}Hz$; 
(b): $s=-0.031$, $t=1.25$, $\epsilon_2=6\epsilon_0$, $\sigma_2=0.02S/m$, 
$\triangle \epsilon_1=1000\epsilon_0$ and $f_1=2.65\times 10^{5}Hz$; 
(c): $s=2.78$, $t=1.053$, $\epsilon_2=78\epsilon_0$, $\sigma_2=2\times10^{-3}S/m$, 
$\triangle \epsilon_1=2500\epsilon_0$ and $f_1=1.9\times 10^{6}Hz$.}
\end{figure}

\begin{figure}[h]
\caption{The characteristic frequencies $f_{b1}=f_{c1}(s_1)$, $f_{b2}=f_{c2}(s_1)$, $f_{b3}=f_{c1}(s_2)$, 
 $f_{b4}=f_{c2}(s_2)$, $f_{b5}=f_{c1}(s_3)$, $f_{b6}=f_{c2}(s_3)$,and the dispersion strengths $\triangle b_1=V_1A(s_1)$,
 $\triangle b_2=V_1B(s_1)$, $\triangle b_3=V_2A(s_2)$,
 $\triangle b_4=V_2B(s_2)$, $\triangle b_5=V_3A(s_3)$ and 
 $\triangle b_6=V_3B(s_3)$ for CMF ($b_z$) versus $L_z$.
The parameters are same as those in Fig.1.}
\end{figure}

\begin{figure}[h]
\caption{Effective dielectric response ($\epsilon_e/\epsilon_0$ and $\sigma_e$) (a) and 
normalized dielectrophoresis (NDEP) and electrorotation (NER)  (b) versus the frequency $f$ 
for the parameters 
in Fig.1(a). In the insert of Fig.3(b), the crossover frequency $f_{cross}$ is 
plotted against $L_z$.}
\end{figure}

\begin{figure}[h]
\caption{Same as Fig.3, but for the parameters in Fig.1(b).}
\end{figure}

\begin{figure}[h]
\caption{Same as Fig.3, but for the parameters in Fig.1(c). Note that two shape-dependent 
crossover frequencies are predicted.}
\end{figure}

\begin{figure}[h]
\caption{Contributions of 
dielectric response $\epsilon_{ei}/\epsilon_0\equiv \triangle \epsilon_{ei}/[1+(f/f_{ei})^2]$ $(i=1,4)$
 to the dieletric spectra $(\epsilon_e-\epsilon_H)/\epsilon_0$  (a), and 
 contributions of i-th strengths $NER_i\equiv \triangle b_i f f_{bi}/(f_{bi}^2+f^2)$  to $NER$  (b) 
 versus $f$ for $L_z=0.05$. Other 
 parameters are same as those in Fig.1(c). Note that the effective dieletric spectra and the $NER$ spectra
 mainly exhibit two-step rapid decrease and  two dominant peaks respectively, 
 although four and  six characteristic frequencies are 
analytically predicted.}
 \end{figure}
 
 \begin{figure}[h]
\caption{$(\epsilon_e-\epsilon_H)/\epsilon_0$ and $NER$ versus $f$ for $L_z=0.95$
and various $\triangle\epsilon_1$. Other parameters are same as those in Fig.1.}
\end{figure}

\begin{figure}[h]
\caption{$(\epsilon_e-\epsilon_H)/\epsilon_0$ and $NER$ versus $f$ for $L_z=0.95$ 
and various $f_1$. Other parameters are same as those in Fig.1.}
\end{figure}

\begin{figure}[h]
\caption{$NER$ versus $f$ with Eq.(16) and Eq.(17) for $L_z=0.05$. 
Other parameters are same as those in Fig.5.
Note that Eq.(16) is independent of $p$.}
\end{figure}

\begin{figure}[h]
\caption{ $(\epsilon_e-\epsilon_H)/\epsilon_0$ and 
$NER$ versus $f$ for $p=0.2$, $L_z=0.95$ and 
$\triangle \epsilon_1=1\times 10^4\epsilon_0$. Other parameters are  same as those in Fig.1(c).
There are four characteristic frequencies, at which the dielectric spectra exhibit 
rapid decrease.}
\end{figure}

\newpage
\centerline{\epsfig{file=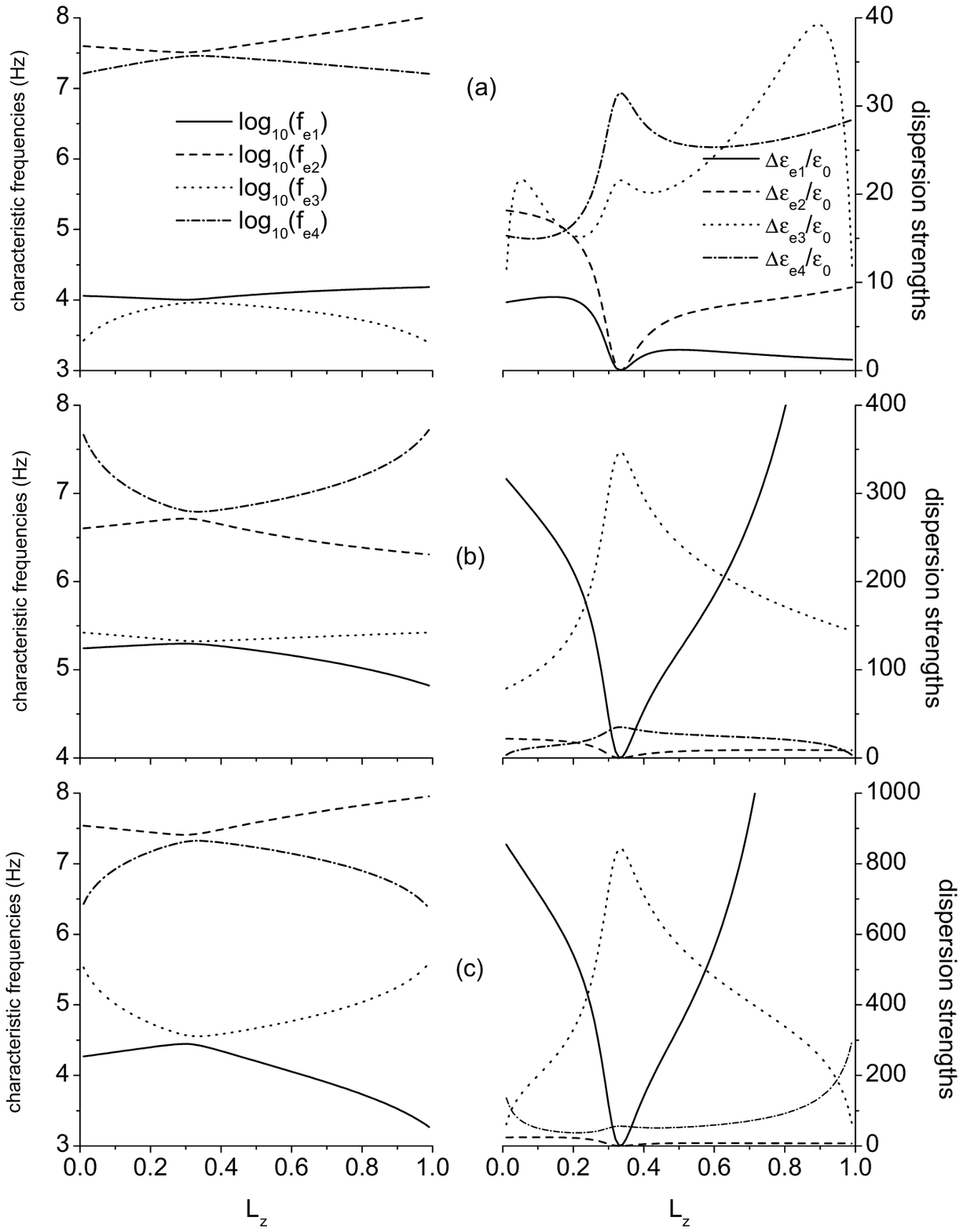,width=400pt}}
\centerline{Fig.1./Gao, Huang, and Yu}

\newpage
\centerline{\epsfig{file=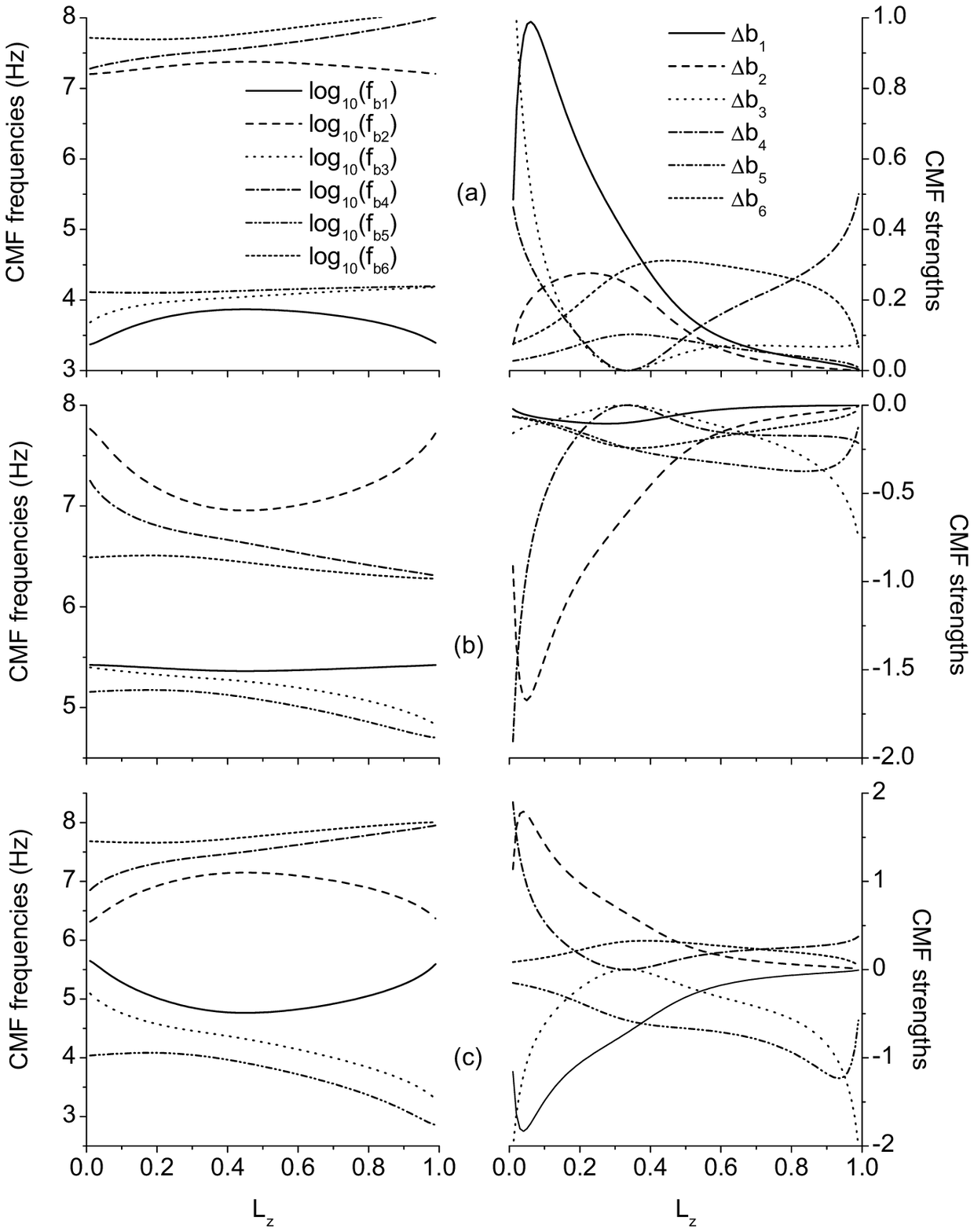,width=400pt}}
\centerline{Fig.2./Gao, Huang, and Yu}

\newpage
\centerline{\epsfig{file=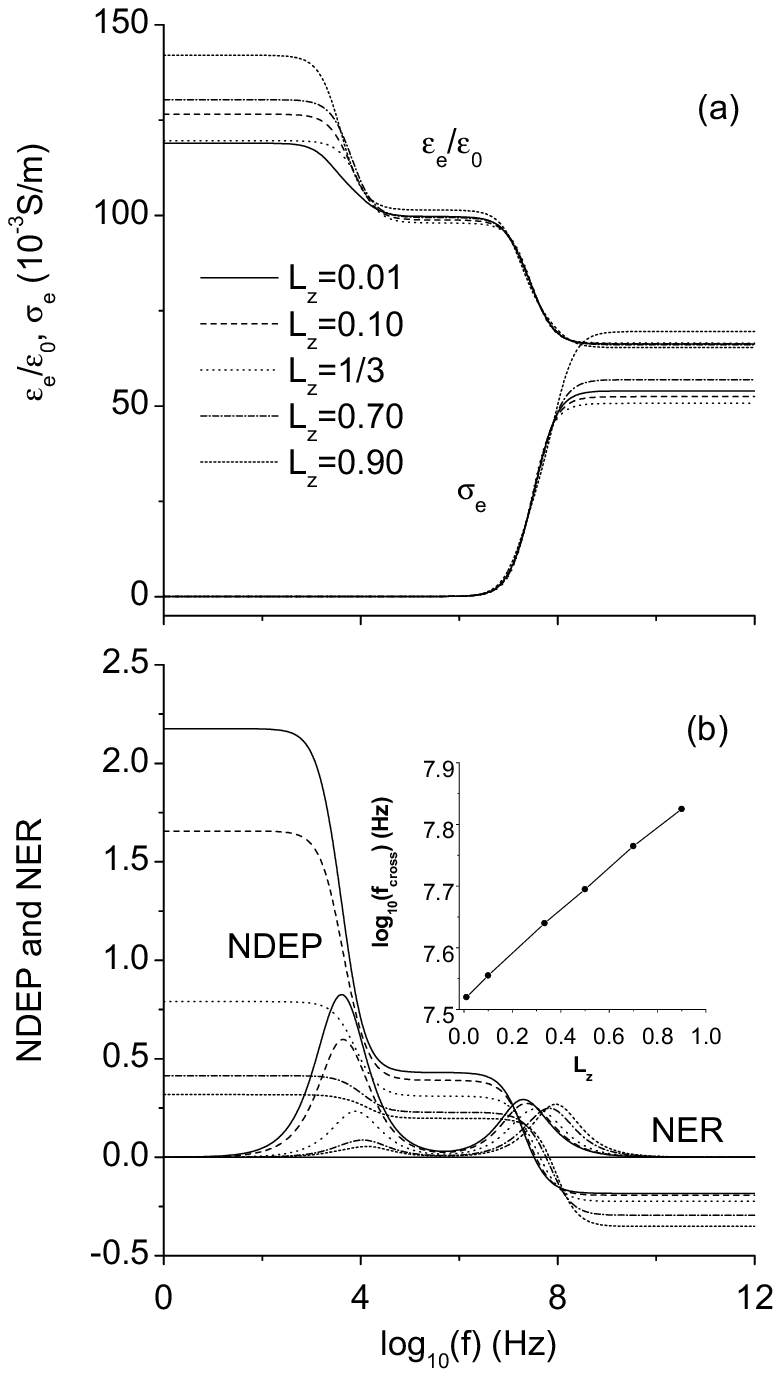,width=400pt}}
\centerline{Fig.3./Gao, Huang, and Yu}

\newpage
\centerline{\epsfig{file=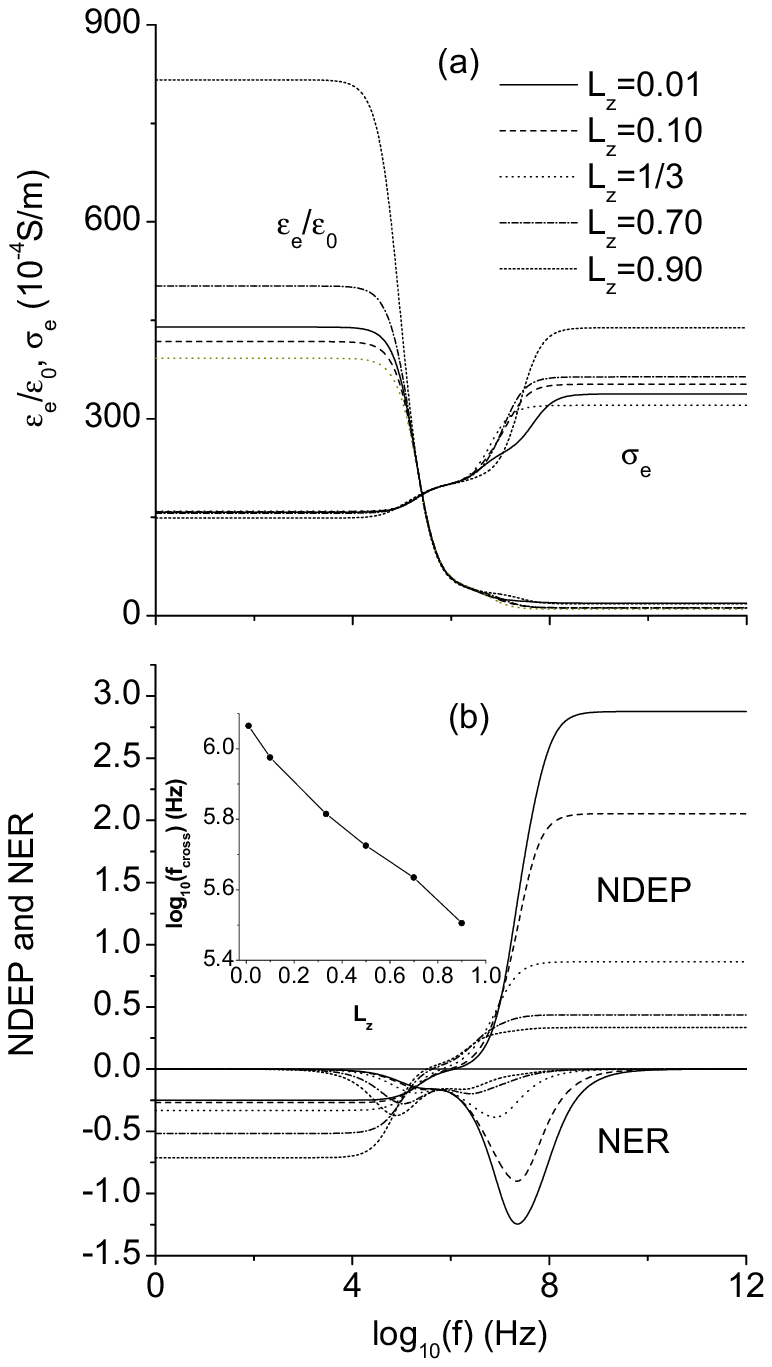,width=400pt}}
\centerline{Fig.4./Gao, Huang, and Yu}

\newpage
\centerline{\epsfig{file=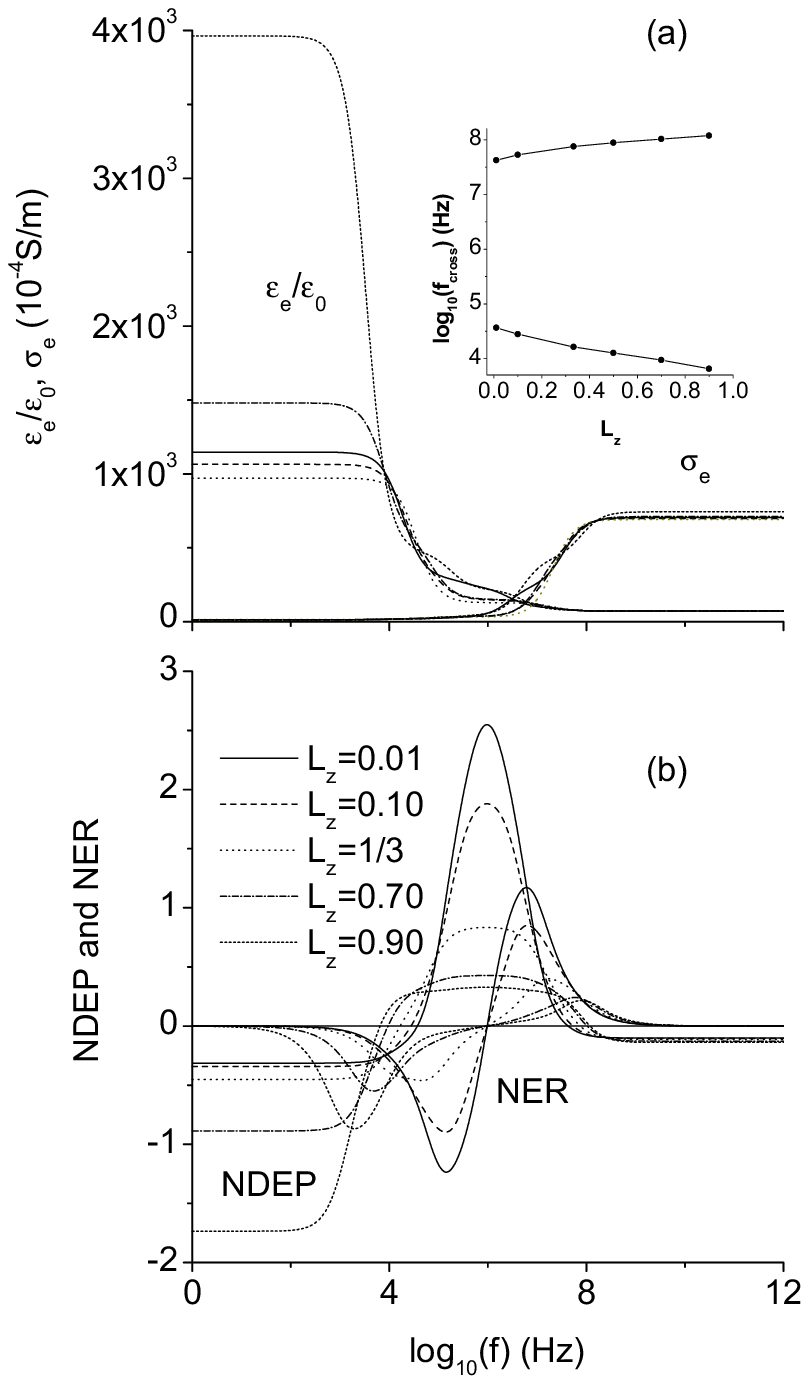,width=400pt}}
\centerline{Fig.5./Gao, Huang, and Yu}

\newpage
\centerline{\epsfig{file=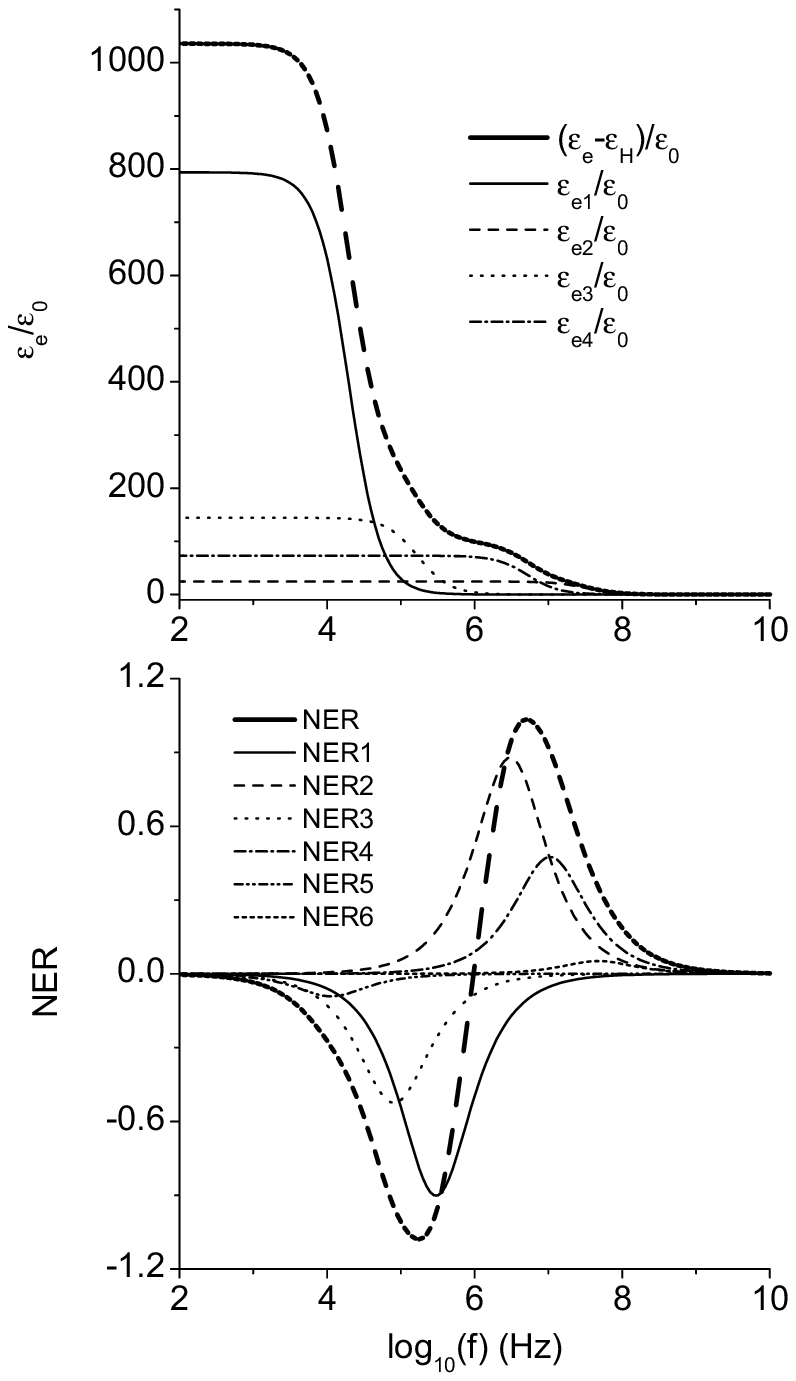,width=400pt}}
\centerline{Fig.6./Gao, Huang, and Yu}

\newpage
\centerline{\epsfig{file=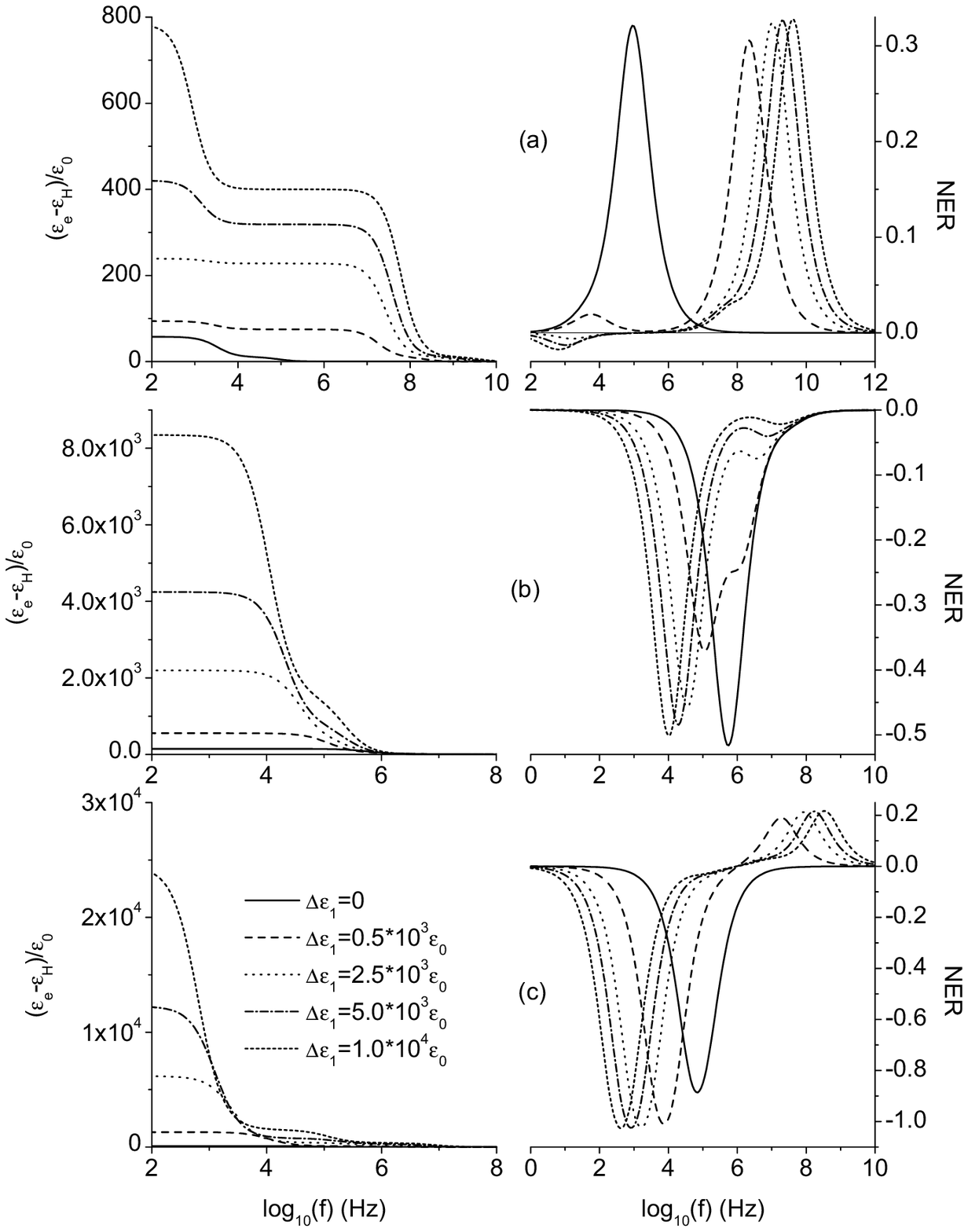,width=400pt}}
\centerline{Fig.7./Gao, Huang, and Yu}

\newpage
\centerline{\epsfig{file=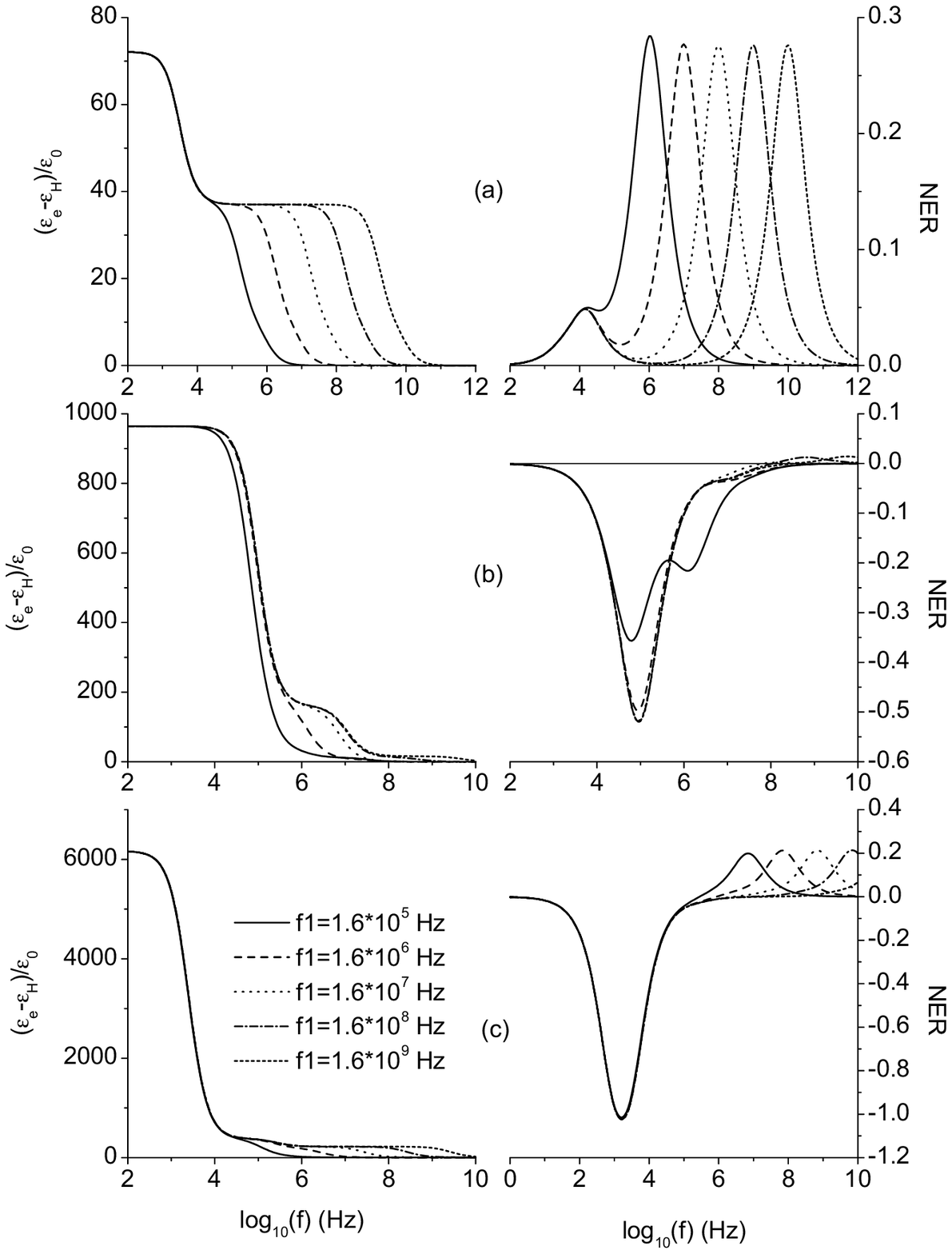,width=400pt}}
\centerline{Fig.8./Gao, Huang, and Yu}

\newpage
\centerline{\epsfig{file=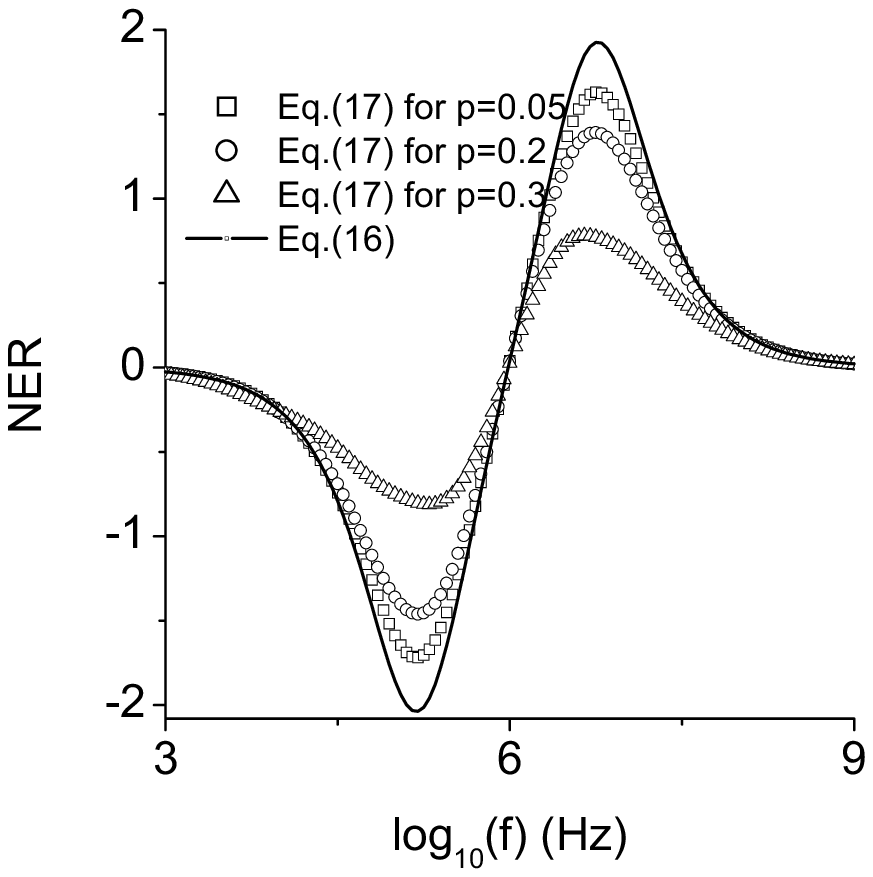,width=400pt}}
\centerline{Fig.9./Gao, Huang, and Yu}

\newpage
\centerline{\epsfig{file=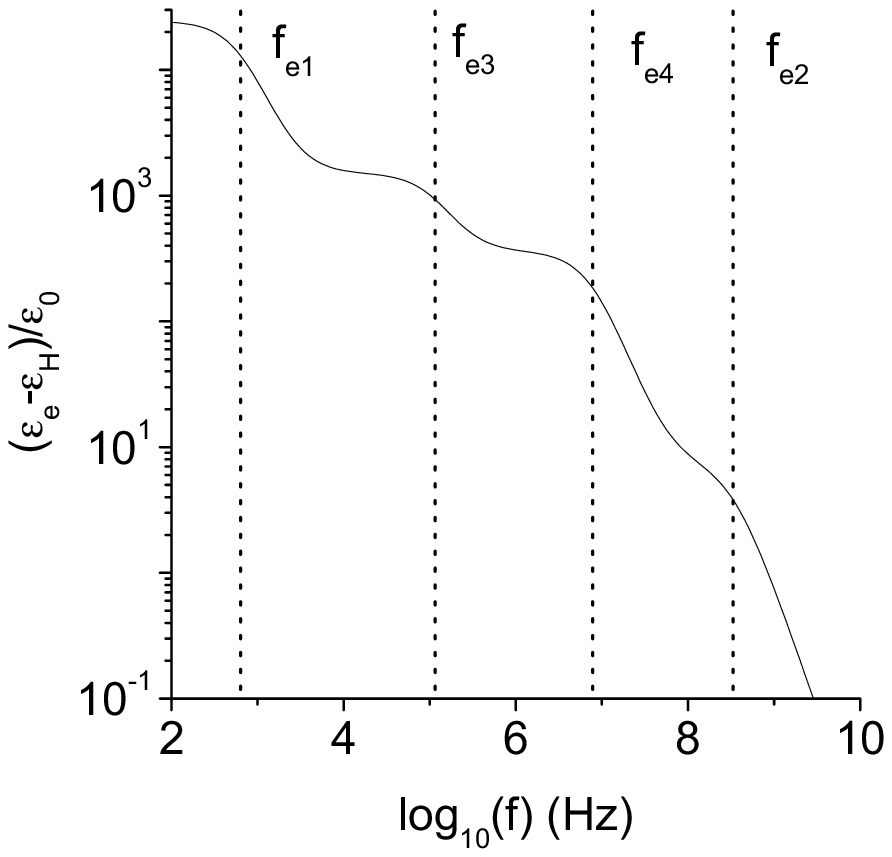,width=400pt}}
\centerline{Fig.10./Gao, Huang, and Yu}

\end{document}